\documentclass[12pt]{article}

\usepackage{times}
\usepackage{graphicx}
\usepackage{siunitx}
\usepackage{bm}

\usepackage{color}

\usepackage{gensymb}
\usepackage{lineno}

\newenvironment{bigabstract}{%
\begin{quote} \bf}
{\end{quote}}

\title{Observation of spin-momentum locked surface states in amorphous $\mathbf{Bi_{2}Se_{3}}$}

\author
{Paul Corbae,$^{1,3\ast}$ Samuel Ciocys,$^{2,3\ast}$ Daniel Varjas,$^{7,8}$ Ellis Kennedy,$^{1,5}$\\ Steven Zeltmann,$^{1,5}$ Manel Molina-Ruiz,$^{2}$ Sin\'{e}ad M. Griffin,$^{3,6}$\\ Chris Jozwiak,$^{4}$ Zhanghui Chen,$^{3}$ Lin-Wang Wang,$^{3}$ Andrew M. Minor,$^{1,5}$\\ Mary Scott,$^{1,5}$ Adolfo G. Grushin,$^{9}$ Alessandra Lanzara,$^{2,3}$ and Frances Hellman$^{1,2,3}$\\
\\
\normalsize{$^{1}$Department of Materials Science, University of California,}\\
\normalsize{Berkeley, California, 94720, USA}\\
\normalsize{$^{2}$Department of Physics, University of California,}\\
\normalsize{Berkeley, California, 94720, USA}\\
\normalsize{$^{3}$Materials Science Division, Lawrence Berkeley National Laboratory,}\\
\normalsize{Berkeley, California, 94720, USA}\\
\normalsize{$^{4}$
 Advanced Light Source, Lawrence Berkeley National Laboratory,}\\
\normalsize{Berkeley, California, 94720, USA}\\
\normalsize{$^{5}$
 National Center for Electron Microscopy,}\\
\normalsize{Molecular Foundry, Lawrence Berkeley National Laboratory,}\\
\normalsize{Berkeley, California, 94720, USA}\\
\normalsize{$^{6}$
 Molecular Foundry, Lawrence Berkeley National Laboratory,}\\
\normalsize{Berkeley, California, 94720, USA}\\
\normalsize{$^{7}$Department of Physics, Stockholm University, AlbaNova University Center,}\\
\normalsize{106 91 Stockholm, Sweden}\\
\normalsize{$^{8}$QuTech and Kavli Institute of NanoScience, Delft University
of Technology,}\\
\normalsize{Delft, The Netherlands}\\
\normalsize{$^{9}$Univ. Grenoble Alpes, CNRS, Grenoble INP, Institut N\'eel,}\\
\normalsize{38000 Grenoble, France}\\
\\
\normalsize{$^\ast$To whom correspondence should be addressed; E-mail: pcorbae@berkeley.edu}
}

\date{}

\begin{document} 


\baselineskip24pt


\maketitle

\begin{bigabstract}

Crystalline symmetries have played a central role in the identification and understanding of quantum materials. The use of symmetry indicators and band representations have enabled a classification scheme for crystalline topological materials, leading to large scale topological materials discovery. 
In this work we investigate whether an amorphous analog of a well known three-dimensional strong topological insulator, which lies beyond this classification due to the lack of long-range structural order, has topological properties in the solid state.
We study amorphous Bi$_2$Se$_3$ thin films, which show metallic behavior and high bulk resistance. The observed low field magnetoresistance due to weak antilocalization demonstrates a significant number of two-dimensional surface conduction channels. Our angle-resolved photoemission spectroscopy data is consistent with a dispersive two-dimensional surface state that crosses the bulk gap. 
Spin resolved photoemission spectroscopy shows this state has an anti-symmetric spin texture, confirming the existence of spin-momentum locked surface states.
We discuss these experimental results in light of a theoretical photoemission spectra obtained with an amorphous tight-binding topological insulator model, contrasting it with alternative explanations. 
The discovery of spin-momentum locked surface states in amorphous materials opens a new avenue to characterize amorphous matter, and triggers the search for an overlooked subset of quantum materials outside of current classification schemes, as a novel route to develop promising scalable quantum devices.
\end{bigabstract}

\section*{Introduction}

Much of materials science and condensed matter physics has focused on exploiting crystal symmetries to understand physical properties, headlined by topological phases and spontaneously broken symmetries in quantum materials. The unusual properties of topological materials, such as the robustness to disorder and their quantized electromagnetic responses, have prompted extensive efforts to classify crystalline topological matter. Non-magnetic crystalline topological insulators and metals with topological bands close to the Fermi level are relatively abundant, representing $\sim$50$\%$ of all materials \cite{Zhang2019,Vergniory:2019ik,Tang:2019jx}, a number that may increase by including magnetic space groups \cite{Watanabe:2018bc}. To identify topological crystals one asks if the band representations of a particular space group admit a trivial insulator limit compatible with the crystal symmetries; if not, the material is labeled topological. The absence of a crystal lattice places amorphous matter outside this classification, even though it is a subset of materials of comparable size to their crystalline counterpart. This raises the question we have set to answer in this work: is there an amorphous topological insulator in the solid state?

Theoretically, amorphous matter can be topological since there are non-spatial symmetries, such as time-reversal symmetry, that protect topological phases. Topological insulator crystals are robust against disorder; topological states do not rely on a periodic crystal lattice at all. In the presence of time reversal invariant disorder the topological states will remain robust and not localize, unless the disorder closes the bulk energy gap, \cite{RevModPhys.82.3045,RevModPhys.83.1057}. Therefore, amorphous materials, which lack translational symmetry and cannot be understood in the context of Bloch states, can still present topological properties. Specifically, electrons in a lattice of randomly distributed atoms with strongly disordered electron hoppings---so strong that no memory of a lattice can be used to label the sites---can present topologically protected edge states and quantized Hall conductivity, hallmarks of topological insulators~\cite{Agarwala:2017jv,Mansha:2017be,Poy17,Mitchell:2018gr,Huang:2018bx,Huang:2018gu,Bourne:2018jr,Marsal30260,spring2021amorphous}. As a proof of principle, a random array of coupled gyroscopes~\cite{Mitchell:2018gr} was designed to act as a mechanical analogue of an amorphous topological state with protected edge oscillating modes, but there has so far been no experimental realization in a solid state material system. 

In this work, we have grown and characterized thin films of amorphous Bi$_2$Se$_3$. The temperature and field dependent resistance reveals the existence of low dimensional carriers with a reduced bulk contribution. 
Angle-resolved photoemission (ARPES) and spin-resolved ARPES show a two dimensional surface state with strong spin-momentum locking, the spin-polarization changes with ARPES detection angle which is proportional to the plane-wave momentum $k$. In its crystalline form Bi$_2$Se$_3$ is a textbook three-dimensional topological insulator \cite{Zhang:2009ks}. 
We find that amorphous Bi$_2$Se$_3$, despite being strongly disordered and lacking translational invariance, hosts two dimensional spin-momentum locked surface states, while nanocrystalline Bi$_2$Se$_3$ does not. 
By numerically simulating a model for amorphous Bi$_2$Se$_3$ with trivial and topological phases, we show that dispersive spin-locked surface states exist in amorphous matter with strong spin-orbit coupling, allowing us to discuss their origin.

\section*{Results}

High resolution transmission electron microscopy (HRTEM) on amorphous Bi$_2$Se$_3$ thin films, Fig. 1(a), shows no signs of crystalline order or even precursor lattice fringes which would have suggested incipient nanocrystals were starting to form. The diffraction pattern in the Fig. 1(a) inset is typical of amorphous materials with well-defined nearest neighbor coordination and inter-atomic distance. There is a diffuse but well-defined inner ring corresponding to the short-range ordering of nearest neighbors. To further ensure we are not probing nanocrystalline regions, scanning nanodiffraction was performed and shown in Fig. 1(b). Four select beam spots each separated by \SI{5}{nm} show amorphous speckle \cite{Treacy_2005} and no signs of Bragg peaks. The speckle visible in the scanning nanodiffraction images from Bi$_2$Se$_3$ is associated with local ordering and orientations of clusters of atoms, which correspond to near- and off-Bragg conditions. The diffracted intensity originates from nanoscale volumes within the sample. The speckle observed across the Bi$_2$Se$_3$ diffraction images is highly uniform, which is indicative of many randomly oriented nanoscale clusters of atoms with short range ordering in the amorphous structure. The diffracted intensity versus scattering vector $k$ for eight different regions is shown in Fig. 1(c), with a single peak corresponding to the diffuse diffraction ring. The nearest neighbor spacing set by the ring is \SI{2.39}{\angstrom}, compared to \SI{3.2}{\angstrom} in crystalline Bi$_2$Se$_3$ \cite{Friedensen:2018tem}. To further investigate the structure, fluctuation electron microscopy (FEM) was used to determine the variance of the diffracted intensity, from which medium range order (MRO) can be extracted and compared to that of nanocrystalline Bi$_2$Se$_3$. FEM is a scanning nanodiffraction technique that probes MRO in amorphous materials through statistical analysis of the variance in diffracted intensity as a function of scattering vector across many diffraction patterns. Fig. 1(d) presents the normalized variance of the diffracted intensity for amorphous Bi$_2$Se$_3$ and nanocrystalline Bi$_2$Se$_3$. The variance in diffracted intensity is a measure of the squared deviation (or fluctuations) in the intensity from the mean intensity at a specific scattering vector. The variance is normalized by dividing by the mean intensity for the specific scattering vector \cite{VOYLES2002147}. Bragg scattering from crystalline regions induces large variations in intensity compared to scattering from amorphous materials. The nanocrystalline sample shows a very large variance with multiple strong peaks due to crystalline order while the amorphous sample does not, providing clear evidence the amorphous samples are indeed amorphous. The electron diffraction scan shows a diffuse diffraction ring over the entire sample; together with the HRTEM images which show no sign of nanocrystals or precursor lattice fringes, these prove the amorphous nature of the film. The samples require selenium capping and subsequent decapping in order to preserve the surface for ARPES. To verify that the decapping process does not generate nanocrystalline regions, Fig. 1(e) displays a XRD $2\theta$ scan for amorphous Bi$_2$Se$_3$, showing a strong substrate peak and a low angle bump typical of amorphous materials which lack long range order but still maintain a well defined interatomic spacing. Electron diffraction confirms the decapped film is still amorphous. The Raman spectrum, Fig. 1(f), shows one broad peak between \SIlist{135;174}{cm^{-1}}. As the laser power increases two peaks can be resolved, which correspond to the bulk E$_{g}^{2}$ and A$_{1g}^{2}$ vibrational modes, respectively. The A$_{1g}^{1}$ van der Waals mode at $\sim$\SI{72}{cm^{-1}}, which is created by the layered structure of the crystal, is absent in our samples. Instead, we observe a peak at \SI{238}{cm^{-1}} not present in crystalline Bi$_2$Se$_3$, which we attribute to amorphous Se-Se bonding as seen in Se films \cite{LUCOVSKY1967113}. The Raman peaks broaden compared to the crystalline system; the full width half maximum of the E$_{g}^{2}$ mode is \SI{23.7}{cm^{-1}} compared to \SI{8.0}{cm^{-1}}~\cite{Zhang:2011raman}. Additionally, EDS maps confirm there is no clustering of Bi and Se in our films and show minimal spatial variations, further confirming that the films do not contain clusters or nanocrystals.
These results show that our samples are amorphous and, while lacking a layered structure with a van der Waals gap, have a local bonding environment similar to the crystalline phase. Moreover, the decapping process important for the following sections does not induce crystallization.

Figure 2(a,b) shows the temperature dependent transport data for different thicknesses in amorphous Bi$_2$Se$_3$, as well as that for the nanocrystalline Bi$_2$Se$_3$ (Fig. 2(e)). The resistivity, $\rho\left(T\right)$, is shown in Fig. 2(a). The $\rho\left(T\right)$ values ($\sim$70-140m$\Omega \cdot$cm) are larger than the crystalline system ($\sim$1-2m$\Omega \cdot$cm \cite{PhysRevB.84.073109}). The amorphous system also demonstrates a much weaker $T$ dependence than the crystalline counterpart \cite{Analytis2010}. As is typical in amorphous metals, the carrier mean free path is determined more by disorder-driven localization than phonon interactions, leading to a largely temperature independent resistivity \cite{RevModPhys.75.1085}. Moreover the high $\rho$ and the weak temperature dependence is inconsistent with either a purely metallic or purely insulating material, and suggestive of a metallic surface on a localized bulk state. Due to the potential metallic surface, we consider the resistance in Fig. 2(b), $R\left(T\right)=\rho\left(T\right) \cdot L/wt$, where $t$ is the film thickness. Again, the $R\left(T\right)$ values are greatly increased compared to the crystal ($\sim$5-250m$\Omega$ for similar thicknesses \cite{PhysRevB.84.073109}). While $R\left(T\right)$ is largely temperature independent, the thicker films show a more pronounced bulk behavior at high temperature \cite{Xu2014}, seen in the inset of Fig. 2(b).
$R(T)$ for each thickness saturates at low temperature; this is in contrast to crystalline Bi$_2$Se$_3$ which shows a low temperature upswing. The low temperature values range from $\sim \frac{h}{3e^{2}}$ to $\frac{h}{e^2}$, similar to bulk insulating topological insulators when gated to bring the Fermi level into the bulk gap \cite{Kim2013,PhysRevLett.123.207701,Xu2016,PhysRevB.99.195302,PhysRevMaterials.1.054202}. The low temperature $R\left(T\right)$ values vary with thickness, possibly a result of small variations in composition between sample thicknesses ($<1$ at$\%$ deviation) or thickness dependent defect formation \cite{PhysRevB.84.073109}.
The effect of composition needs to be explored further. Transport and ARPES results were reproducible on different samples. The $R\left(T\right)$ data was fit using a two channel conductance model \cite{doi:10.1063/1.4719196} represented by the dashed curves in Fig. 2(b). The model requires parallel contributions from an insulating, variable range hopping bulk and metallic surface states, and provides an overall good fit to the data. Below around \SI{150}{K}, the surface contribution dominates for films of all thicknesses. These results indicate that the surface state contribution to conduction is metallic and dominant over a large temperature range for the amorphous samples. The conductivity in nanocrystalline Bi$_2$Se$_3$ (shown in Fig. 2(e)) drops over the entire temperature range and does not display any metallic behavior with temperature.

The magnetoconductance (MC) provides another means to probe the transport. Fig. 2(c) shows MC data for a \SI{140}{nm} film, revealing a sharp decrease in the low field $\Delta G$ ($<$ 2 T) at low temperatures which is typical of weak anti-localization (WAL), consistent with a metallic surface state in amorphous Bi$_2$Se$_3$ \cite{PhysRevLett.105.176602}. This is in contrast to other non-magnetic amorphous systems which are topologically trivial and show a MC increase due to weak localization \cite{Korzhovska2020}. The case of positive MC is likely due to a diminished spin-orbit coupling (SOC) effect where disorder closes the mobility gap or a local environment that does not produce topological states. In amorphous  Bi$_2$Se$_3$ the nearest neighbor distance is smaller than in the crystal, acting similar to pressure, which has been shown to lead to a topological gap \cite{Bahramy2012bitei}. In amorphous Bi$_2$Se$_3$, due to strong SOC, backscattering is suppressed when a field is absent and time reversal symmetry is present. When time-reversal symmetry is broken with the application of a magnetic field, backscattering increases leading to a positive MR. The magnetoconductance can be fit with the standard Hikami-Larkin-Nagaoka (HLN) formula for WAL \cite{10.1143/PTP.63.707}, \(\Delta G\left(B\right) = \alpha \frac{e^2}{\pi h}\left[\Psi\left(\frac{\hbar}{4eBl_{\phi}^{2}}+\frac{1}{2}\right)-\ln\left(\frac{\hbar}{4eBl_{\phi}^{2}}\right)\right]\)
where $\Psi$ is the Digamma function, $B$ is the out-of-plane field, $l_{\phi}$ (the phase coherence length) and $\alpha$ are used as fitting parameters. According to this model, each conductance channel with a $\pi$ Berry phase should contribute an $\alpha=-1/2$ factor to $\Delta G$ \cite{PhysRevLett.97.146805}. 
Fitting our low field data, Fig. 2(d), at \SI{2}{K} gives a value of  $\alpha=-0.81$, suggesting we have two decoupled surface states 
\cite{PhysRevLett.113.026801}. At \SI{20}{K}, $\alpha=-0.51$ suggesting the surface states are coupled to a bulk state, causing the entire film to act effectively as one channel, as seen in crystalline Bi$_2$Se$_3$ from \SIrange{2}{100}{nm} \cite{PhysRevB.84.233101,Liao2017}.
As the temperature increases the WAL contribution is diminished. 
Based on Hall measurements, the two-dimensional carrier density is $n_{2D}=2.8 \times 10^{14}$ cm$^{-2}$ and the three-dimensional carrier density is $n_{3D}=1.9 \times 10^{19}$ cm$^{-3}$, leading to a mobility of 21.8 cm$^{2}/$Vs. According to the Ioffe-Regel criterion amorphous Bi$_2$Se$_3$ has $k_{F}l\sim1$ \cite{BRAHLEK201554} and have similar $\mu$,$n_{3D}$ values reported in the bulk insulating BiSbTeSe solid solution \cite{PhysRevB.84.165311}. Additionally, the calculated mean free path at \SI{2}{K} is $\sim \SI{1}{nm}$. This $n_{3D}$ likely places $E_F$ into the conduction band (seen in ARPES presented below), although the depth depends on the effective mass \cite{BRAHLEK201554}. Since our system is amorphous the bulk carriers are expected to be localized and provide little contribution to the transport, leading to the observed high $\rho$. The observed behavior in the amorphous Bi$_2$Se$_3$ sheet resistance and MR is a result of metallic surface states that dominates over a wide range of temperatures.

Amorphous materials are not expected to have any electronic states with well-defined momenta, but are nonetheless known to support metallic conduction and superconductivity.
However, since the nearest neighbor distance is well defined (inset Fig. 1(a)), there exists a good reciprocal length scale. 
If sharp spectral features are observed, which is the case in our ARPES measurements, there exist states with good momentum quantum numbers since this corresponds to the overlap of the electronic wave-functions with plane waves of well defined $k$, modulated by matrix elements. The coordinates $\theta$ and $\phi$ are experimentally measured and refer to the respective angles of photoemission from the sample surface. The plane wave components $k_x$ and $k_y$ are proportional to $\theta$ and $\phi$ at small angles, where $\theta$ is the azimuthal angle and $\phi$ is the polar angle. In our work we refer to spin-momentum locking as the spin asymmetry around zero angle since at small angles, the momentum of the plane wave and the detection angle are proportional.

To interpret our experimental data and determine if it is consistent with a topological bulk we developed a numerical model that realizes a Dirac-like state in the absence of crystalline symmetry. Motivated by the similarity of local environments between the crystalline and amorphous Bi$_2$Se$_3$ found in Fig. 1, we use an amorphous variant of the three-dimensional four-band (spin-1/2 x 2-orbital) BHZ model~\cite{Zhang:2009ks,Agarwala:2017jv}. From the model we numerically obtain a spin resolved spectral function, shown in Fig. 3(a,b), for both the trivial and topological phase of amorphous Bi$_{2}$Se$_{3}$, respectively. While inversion is expected to be an average bulk symmetry in amorphous solids, it will be broken by the surface, allowing us to include a surface onsite potential that breaks this symmetry. As for the crystal surface states, this term spin-splits trivial surface states at $E_{F}$. This surface potential depends on the details of the surface termination (such as dangling bonds, Se vacancies, or surface reconstruction~\cite{PhysRevB.81.041405,PhysRevLett.103.146401}) and can tune the Dirac point to arbitrary binding energies~\cite{Bahramy2012, Bianchi2010, PhysRevLett.107.086802}. However, it does not affect the bulk topological properties. In the trivial phase we observe spin-split states symmetrica round $\phi=0$ above the gap, while in the topological state a Dirac cone pinned to $\phi=0$ is visible, guaranteed by time-reversal symmetry, spanning the bulk gap (see Fig. 3(a,b)).

Fig. 3(c) displays the raw ARPES spectrum as a function of energy and emission angle $\phi$ at a specific $\theta$, a momentum space slice that intersects $\Gamma \left(\phi=\SI{0}{\degree}\right)$. The dispersion revealed here in amorphous Bi$_2$Se$_3$ marks the first observation of an amorphous band structure with sharp, momentum-dependent features. Notably, the dispersion exhibits two vertical features at the Fermi level crossing the bulk gap.
The raw spectrum reveals an intensity peak near $E_F$ starting at \SI{-0.2}{eV} and a sharp rise in intensity below \SI{-0.5}{eV}. The increased intensity of the surface states near $E_F$ may be due to photoemission enhancement from the less-visible bulk conduction band. The increased intensity below \SI{-0.5}{eV} coincides with a less-dispersive band which is most likely the bulk valence band. The exact bottom of the conduction band and top of the valence band is obscured in the ARPES spectra due to intrinsic broadening. However, using angle-integrated photoemission, we can roughly estimate the band gap to be $\sim$\SI{350}{meV}, consistent with the calculated DOS from amorphous structures using ab-initio molecular dynamics (\SI{299}{meV}). 

Fig. 3(d) presents the experimental in-plane Fermi surface in amorphous Bi$_2$Se$_3$. The annular Fermi surface is consistent with crystalline Bi$_2$Se$_3$, where the dispersion associated surface state in Fig. 3(c)  produces a ring at the Fermi surface. The dot product of the $p$-orbital axis and the experimental coordinates is well defined and should lead to a similar scenario as seen in the crystalline case in which $p$-polarized light couples asymmetrical across $k=0$, leading to the observed orbital effect in the Fermi surface. To confirm that these states are localized to the surface, in Fig. 3(e)  we show the photon energy plotted versus emission angle $\phi$. 
Due to conservation of energy, photon energy ($h\nu$) and $k^2$ of the photoemitted electron are nearly-proportional for large $h\nu$, and related by $\hbar^2k^2/2m = h\nu - W - E_b$ where $W$ is the work function of the material and $E_b$ is the binding energy. In the plane $(k^2,\phi)$ the states are nearly independent of photon energy (red lines in Fig. 3(e)). For a 3D amorphous system (or even a  3D polycrystalline system), bulk states must be spherically symmetric and independent of $\phi$ due to the absence of an average preferred direction. Therefore the strong $\phi$-dependence and $h\nu$-independence suggests the electrons are not from the bulk and instead originate from surface states. These observations motivate us to interpret these states as two-dimensional surface states. It is important to note that there exist significant density of states at the Fermi energy associated with the 2D surface states, confirming a two-dimensional transport channel as determined by our magnetoresistance measurements. 

The presence of strong SOC added to broken inversion symmetry at the surface in our system should lead to a spin texture. Fig. 4 shows spin-resolved angle-resolved photoemission spectroscopy taken with \SI{11}{eV} photons. We observe an anti-symmetric spin-polarization, the first observation in an amorphous system to the best of our knowledge.
The spin-polarized energy distribution curves (EDCs) with $p$-polarized light are shown in Fig. 4(a) at $\phi=$\SIlist{-6;0;6}{\degree}. The spin-polarization is measured by the relative difference between spin-up and spin-down photoelectrons weighted by the Sherman function ($S$) of the detector in the form $P_{y} = S*\left(I_{\uparrow} -I_{\downarrow}\right)/\left(I_{\uparrow} + I_{\downarrow}\right)$. The most evident feature from the three spin-polarized EDCs is the large positive polarization between \num{-0.6} and \SI{0.0}{eV} that reaches a maximum of $\sim$50$\%$. This large polarization offset is due to spin-dependent photoemission matrix elements (SMEs) in which SOC leads to selective emission of electrons with a particular spin-state. This is observed in crystalline Bi$_2$Se$_3$ near the upper Dirac cone with similar intensity~\cite{Jozwiak2013}. 

In order to uncover the intrinsic spin texture (i.e. the sign of the polarization) within the SME background we follow a similar background subtraction to Ref.~\cite{Jozwiak2013}. Fig. 4(b) presents the spin polarization as a function of binding energy and $\phi$ after performing the background subtraction. From this spin-polarized map, three ranges of binding energy demonstrate distinct anti-symmetric spin polarizations with respect to $\phi$: $E_F$ to \SI{-0.20}{eV} (region I), \SI{-0.20}{eV} to \SI{-0.55}{eV} (region II), and \SI{-0.55}{eV} to \SI{-0.75}{eV} (region III). The spin polarization has a magnitude of $\pm$15$\%$ and changes sign between these ranges as a function of binding energy. By comparing Fig. 4(b) with the spin-integrated spectrum in Fig. 3(c) we see that region I corresponds to the conduction band, region III corresponds to the valence band, and region II corresponds to the in-gap states. The in-gap states of region II have opposite spin-polarization to both the conduction band and valence band, suggesting that these states are indeed separate features from region I and III and not a consequence of inelastic scattering from region I states or from local variations in composition. The measured spin-polarization matches the expected spin-polarization from our tight-binding model shown in Fig. 3(b) for the topological case with region I representing the spin texture of the trivial Rashba split bulk states near the Fermi level, region II representing the spin texture of the upper Dirac cone of the topological surface state in the bulk band-gap, and region III representing the spin texture of the lower Dirac cone within the bulk valence band. 
We conclude that the two-dimensional surface states form a node around \SI{-0.55}{eV}, and the anti-symmetric spin-resolved spectrum around $\Gamma$ at $E_F$ is associated with trivial states with a large component at the surface stemming from Rashba-type spin-splitting in our system, as seen also in crystalline Bi$_2$Se$_3$ \cite{Jozwiak2016}. 

Electron microscopy in Fig. 1 confirms the structure is amorphous with a shortened interatomic spacing. Fig. 2 shows the carriers in the system are weakly anti-localized, suggesting two-dimensional decoupled metallic conductance channels. ARPES, Fig. 3, shows dispersive surface states that cross the bulk gap. The midgap states are spin polarized shown via SARPES in Fig. 4. Our data is consistent with a spin-momentum locked two-dimensional surface state for which we discuss two main origins: non-topological and topological. In the latter, a topological bulk state in amorphous Bi$_2$Se$_3$ would produce the dispersive, spin-momentum locked surface states we observe. In this interpretation, the topological surface states cross the bulk gap, as in Fig. 3(b), forming a Dirac state similar to the surface states observed in related topological insulator crystals \cite{doi:10.1063/1.3595309}. 
In the amorphous case however, the Dirac point can be hidden underneath the bulk valence band due to a strong surface potential, as discussed above (recall that, although a Dirac node crossing is a protected feature of topological insulators, the binding energy of the node is not). Additionally, from our tight-binding model, we expect trivial spin-split states of Rashba-type to develop at the Fermi level, with opposite polarization from the topological surface state, due to broken inversion symmetry at the surface, as observed in Ref. \cite{Jozwiak2016}; this will be discussed below.
For the topological explanation to be viable, we need to discard that the amorphous spectrum is a result of averaging over nanocrystalline domains. Fig. 3(f) shows the resulting spectrum from nanocrystalline Bi$_2$Se$_3$ in which averaging over randomly rotated domains leads to angle-independent photoemission. Our spectra is also distinct from the crystalline spectrum \cite{Jozwiak2016}, where the amorphous Bi$_2$Se$_3$ spectral features extend further in $\phi$ than the crystalline case.

Although the topological origin of the surface state is consistent with our data, it is important to discuss non-topological origins like two-dimensional Rashba states, and features that are not captured by our model.
Spin resolved ARPES shows an anti-symmetric spin-polarization on either side of the nodal region observed at \SI{-0.55}{eV}. In the topological scenario these would be attributed to a spin-polarized two-dimensional topological surface state.  Another possible explanation is that the lack of inversion symmetry at the surface gives rise to trivial Rashba states, as in Fig. 3a, and a spin-polarization in the bulk conduction and valence bands. If the spin texture were a result of Rashba splitting, then each near-vertical branch would need to be a single parabolic-like band that returns to the Fermi level. This is because each near-vertical feature has only one spin character. This Rashba interpretation could be plausible if the parabolic dispersion was obscured by the momentum broadening. Indeed, the spectra are broadened by the atomic disorder as well as the presence of vacancies and dangling bonds at the surface, which are a significant source of final state scattering. However, it would also imply a gigantic Rashba momentum offset of $k_0 \approx 0.4$ Å$^{-1}$. For reference, the giant Rashba splitting in bulk BiTeI is demonstrated by $k_0 = 0.051$ Å$^{-1}$ \cite{IBM11}, nearly a factor of 10 smaller. 
While we cannot definitively settle on either explanation, the topological scenario seems simpler as it could explain the spin switching of all regions in Fig. 4, and because it seems that a gigantic Rashba splitting is needed to explain our results otherwise. Regardless of the explanation, our observation of well-defined, spin-momentum locked dispersing surface states opens a new direction to characterize amorphous matter, and to search for new materials with advantageous properties, such as spin-momentum locking.

In the raw spectra ARPES features are perceived as near-vertical, yet the underlying bands need not have infinite electron velocity. They are broadened to the point that the electron velocity is hard to quantify yet we can appreciate that they do have an increased Fermi velocity compared to the crystal. It is known that the surface environment can affect the location of the Dirac node and the curvature of the bands significantly in the crystalline case \cite{PhysRevB.89.125109}. We expect that this is exacerbated in the amorphous case due to the presence of dangling bonds. In fact, amorphous materials have been predicted to experience large band renormalizations that lead to backbending of the dispersions as well as vertical features \cite{Edwards1961,Olson1975,Ryu2021}. Combining these factors, it is reasonable to expect large deviations in the amorphous surface state dispersion from the crystalline case, both in the observed band velocity from renormalizations and burying of the Dirac node in the valence band.
However, the spin-polarization allowed us to argue that states crossing the gap-like region (II) are unrelated to broadening caused by inelastic scattering of conduction band electrons. Specifically, the spin polarization reverses sign going down in energy from the conduction band (I) at $E_\mathrm{F}$ into the the gap-like region (II), reversing again in the valence band (III)
states (i.e. unrelated to broadening caused by inelastic scattering of conduction band electrons) lie within the gap. We do this by revealing the sign reversals of the spin near $E_\mathrm{F}$, in the gap-like region, and again in the valence band.

TEM and Raman data suggests that the typical local structure of the amorphous system are comparable to the crystalline case, indicating that a possible condition to preserve the topological nature of the bulk in the amorphous state is to retain a local ordering similar to that of the crystal \cite{Marsal30260}. However, it should be noted that there is no continuous pathway from the crystalline phase to amorphous phase but instead a discontinuous phase transformation associated with the nucleation of crystalline domains \cite{doi:https://doi.org/10.1002/9783527617968.ch1}. The nucleation process explains why nanocrystalline Bi$_2$Se$_3$ has been shown to be topologically trivial, Fig. 2(e), due to the presence of grain boundaries and other structural defects \cite{PhysRevB.94.165104}. 
The impact of coupling strengths and local field environments can be assessed theoretically via ab-initio calculations to refine the Hamiltonian modeling of amorphous topological materials \cite{Zhang:2009ks}. This approach can be extended to amorphous material systems without topological crystalline counterparts, where local ordering coupled with disorder and strong SOC can mix energy levels to produce a topologically nontrivial electronic structure. 

\section*{Conclusion}

In conclusion, we have found that amorphous Bi$_2$Se$_3$ hosts a dispersing two-dimensional metallic surface state with spin-momentum locking. This experimental observation of spin-momentum locked surface states in an amorphous solid state system highlights that searching for new quantum materials with advantageous properties, such spin-momentum locked states, or topological properties should not be restricted to crystalline solids. 
Our work provides a study of 
an amorphous solid state system 
with chemical specificity and local bonding environments, a system which can be implemented into devices. 
To the best of our knowledge there have been no previous reports of ARPES/SARPES on an amorphous solid. Our results represent the first step towards realizing, in real materials, recently proposed non-crystalline topological phases~\cite{PhysRevLett.123.196401,Marsal30260} that lie outside the known classification schemes for topological crystalline matter \cite{Zhang:2018yqi,Vergniory:2019ik,Tang:2019jx,Watanabe:2018bc,po2017classification} and that may be more robust than their crystalline counterparts~\cite{spring2021amorphous}. We expect our work to motivate an effort to understand topological amorphous matter, enabling materials discovery that can provide a path towards affordable and better implementation into modern thin film processes.

\section*{Methods}

Amorphous Bi$_2$Se$_3$ thin films were thermally evaporated in a UHV chamber with base pressure of $10^{-9}$ Torr. The films were grown at room temperature from high purity ($99.999\%$) elemental Bi and Se single sources. Stoichiometry of the films was confirmed using XPS (X-ray photoelectron spectroscopy), EDS (Energy dispersive X-ray spectroscopy), and RBS (Rutherford backscattering spectroscopy). High resolution TEM and Fluctuation electron microscopy (FEM) were performed on \SI{10}{nm} thick Bi$_2$Se$_3$ films deposited on a \SI{10}{nm} thick SiN window. FEM experiments were performed using an FEI TitanX operated at an acceleration voltage of 200 kV.  Diffraction images were collected on an Orius CCD system with an exposure time of 0.3 seconds and a camera length of 300 mm.  The probe convergence angle was set to 0.51 mrad by adjusting the third condenser lens current, resulting in a probe diameter of 2.2 nm and a probe current of 15.5 pA. Nanodiffraction data were collected as 15-by-15 image stacks (225 total images). Multiple 225-image datasets were collected for both the amorphous and polycrystalline Bi$_2$Se$_3$ for statistical averaging. Each data set covered an area on the film of approximately 77-by-77 nm. The first image from each dataset was excluded to avoid including any potential sample damage or contamination in the data resulting from the parked beam. The central beam was covered using a beam stop and the beam position remained constant across all FEM images for each sample. Variation in peak positions and intensities were negligible across data from different locations on a single film. Imaging conditions were held constant for all data collection to prevent variations in microscope alignment.
The amorphous structure of the film was confirmed with XRD, Raman spectroscopy, and TEM.

The amorphous Bi$_2$Se$_3$ samples $\rho\left(T\right)$ was measured using a four point probe. The samples were grown as a bar using a metal mask onto pre-deposited Au(\SI{5}{nm})/Cr(\SI{2}{nm}) contacts to ensure ohmic contact (shown in Fig. 2(a) inset). Magnetrotransport was measured in the Van der Pauw configuration with samples grown onto pre-deposited Au(\SI{5}{nm})/Cr(\SI{2}{nm}) contacts.

The Hamiltonian used to describe amorphous Bi$_2$Se$_3$ features direction-dependent spin-orbit hoppings set by the normalized hopping vector $\hat{\mathbf{d}}$ and is the sum of onsite and hopping terms
\begin{eqnarray}
H_{\mathrm{onsite}}&=&m\sigma_{0}\tau_{z},\\
H_{\mathrm{hop}}(\hat{\mathbf{d}})&=&it_{1}(\hat{\mathbf{d}}\cdot\boldsymbol{\sigma})\tau_{x}+t_{2}\sigma_{0}\tau_{z}
\end{eqnarray}
where $\sigma_i$ and $\tau_i$ are the spin and orbital Pauli matrices respectively, $m$ sets the splitting between the local $s$ and $p$-like orbitals, $t_1$ is the spin-orbit hopping, and $t_2$ is the normal hopping amplitude.
In the crystalline case this Hamiltonian correctly reproduces key features of the topologically nontrivial bands closest to the Fermi level~\cite{Zhang:2009ks}.
We implement this tight-binding model on large systems of short-range correlated amorphous structures and investigate the topological surface states by calculating spectral functions using the Kernel Polynomial Method~\cite{Weisse:2006go,PhysRevResearch.2.013229}.

We performed ARPES at the Advanced Light Source MAESTRO (7.0.2) and MERLIN (4.0.3) beamlines with photon energies in the range of 65 - 125 eV. ARPES results taken on different samples and at different beamlines produce the same results. The decap procedure does not create any crystalline order in our samples. The spin-resolved spectra were acquired from a high-efficiency and high-resolution spin-resolved time-of-flight (TOF) spectrometer that utilizes the spin-dependent reflection from a magnetic thin-film due to the exchange interaction \cite{doi:10.1063/1.3427223}. The light source for the spin measurements was a Lumeras 11eV Xenon gas-cell laser with 1MHz repetition rate \cite{doi:10.1063/1.4939759}. Synchrotron ARPES measurements and spin-resolved measurements were taken at ~20K and ~75K, respectively. ARPES was analyzed using the PyARPES software package \cite{STANSBURY2020100472}.

Density functional theory (DFT) calculations were performed using the projector augmented wave (PAW) formalism in the Vienna ab initio Simulation Package (VASP) \cite{VASP1, VASP2}. The exchange-correlation potentials were treated in the framework of generalized gradient approximation (GGA) of Perdew-Burke-Ernzerbof (PBE) \cite{PBE1}. Bi (6s, 6p) and Se(4s, 4p) electrons were treated as valence, and their wavefunctions expanded in plane waves to an energy cutoff of 500 eV.  A k-point grid of 3x3x1 with Gamma sampling was used. Spin-orbit coupling was added self-consistently for all density of states calculations. Amorphous structures were generated with ab initio molecular dynamics using VASP.

\bibliography{amorphous.bib}
\bibliographystyle{sc.bst}

\section*{Acknowledgments}
P. C. and S. C. would like to thank E. Parsonnet and D. Rees for their discussions.
A. G. G. is grateful to J. H. Bardarson, S. Ciuchi, S. Fratini, and Q. Marsal for discussions.
D. V. thanks A. Akhmerov, A. Lau and P. Perez Piskunow for discussions.
The project was primarily funded by the U.S. Department of Energy, Office of Science, Office of Basic Energy Sciences, Materials Sciences and Engineering Division under Contract No. DE-AC02-05-CH11231 within the Nonequilibrium Magnetic Materials Program (MSMAG).
The ARPES and SARPES work was supported by Berkeley lab's program on Ultrafast materials sciences, funded by the U.S. Department of Energy, Office of Science, Office of Basic Energy Sciences, Materials Sciences and Engineering Division under Contract No. DE-AC02-05-CH11231.
TEM at the Molecular Foundry was supported by the Office of Science, Office of Basic Energy Sciences, of the U.S. Department of Energy under Contract No. DE-AC02-05CH11231.
Computational resources were provided by the National Energy Research Scientific Computing Center and the Molecular Foundry, DoE Office of Science User Facilities supported by the Office of Science of the U.S.\ Department of Energy under Contract No.\ DE-AC02-05CH11231. The work performed at the Molecular Foundry was supported by the Office of Science, Office of Basic Energy Sciences, of the U.S.\ Department of Energy under the same contract. 
P. C. is supported by the National Science Foundation Graduate Research Fellowship under Grant No. 1752814.
S.C. was supported by the National Science Foundation Graduate Research Fellowship under Grant No. DGE1852814 and DGE1106400.
A. G. G. is supported by the ANR under the grant ANR-18-CE30-0001-01 and the European Union Horizon 2020 research and innovation programme under grant agreement
No. 829044.
D. V. is supported by NWO VIDI grant 680-47-53.
S. E. Z. was supported by the National Science Foundation under STROBE Grant No. DMR 1548924.
\section*{Author contributions}
P. C. and S. C. contributed equally to this work. 
The project was initiated and oversaw by P.C., S.C., A. G. G., A.L., and F. H..
P. C. grew the films.
S.C. performed synchrotron ARPES, S. C. and P. C. performed SARPES measurements, and S.C. performed the data analysis.
P. C. performed transport measurements.
S. Z., E.K., S. C., and P. C. performed TEM.
M. M. R. and P. C. performed Raman measurements.
Z.C. performed the molecular dynamics and S.G. performed the DOS calculations.
A. G. G. and D. V. constructed the tight-binding model, and D. V. performed numerical calculations.
P. C., S. C., A. G. G. and D. V. took part in interpreting the results.
All authors contributed to writing the manuscript.

\clearpage

\begin{figure}[ht!]
\centering
    \includegraphics[width=1\textwidth]{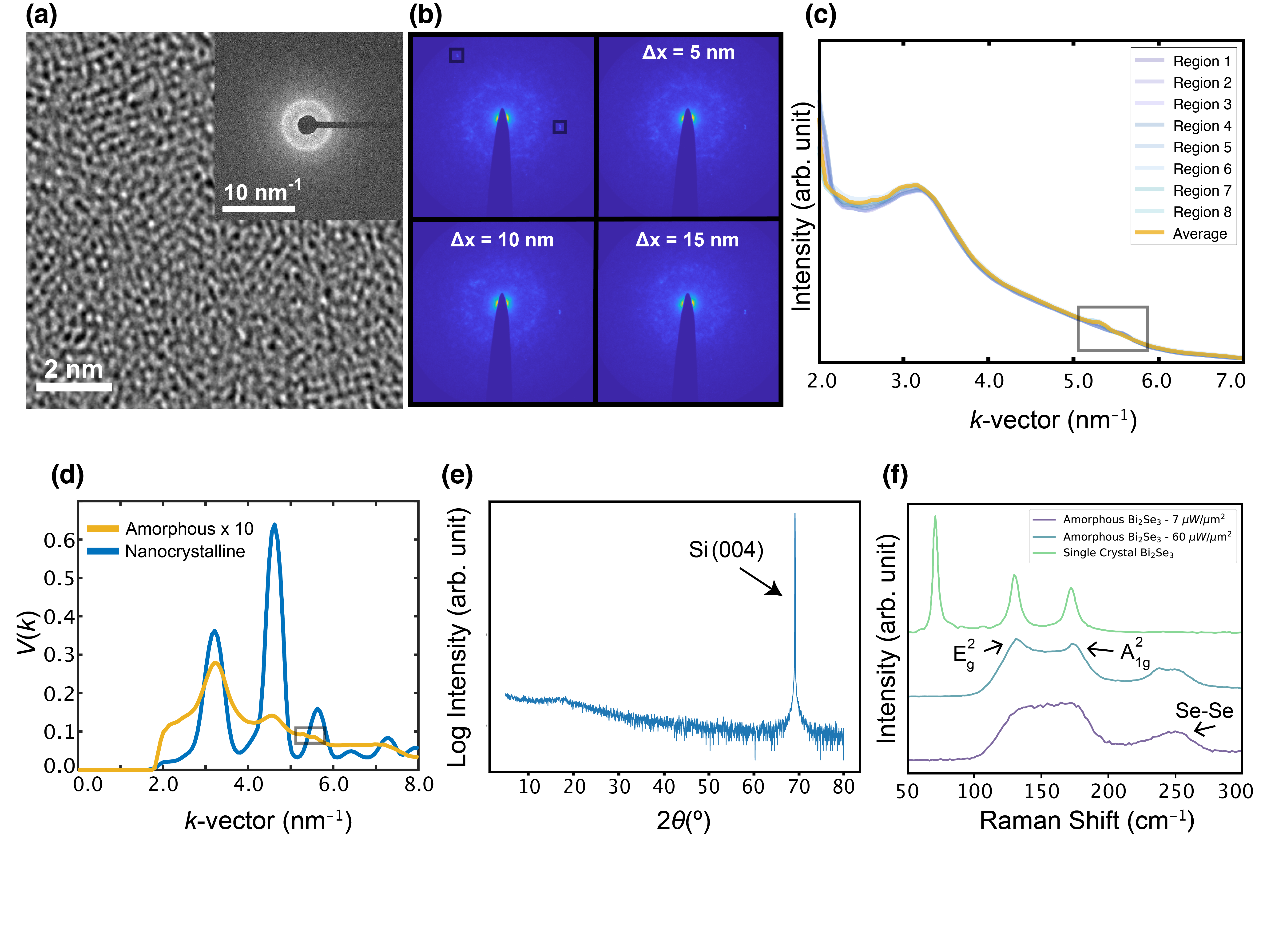}
\end{figure}
\noindent {\bf Fig. 1.} \textbf{Structural and spectral evidence for the amorphous atomic structure of Bi$_2$Se$_3$} \textbf{(a)} HRTEM image. Inset: Diffraction pattern for the amorphous Bi$_2$Se$_3$ films. We observe a diffuse ring due to the amorphous nature of the film corresponding to $\sim\,$\SI{2.4}{\angstrom}. \textbf{(b)} Scanning nanodiffraction patterns taken with a beam spot of \SI{2}{nm} separated by \SI{5}{nm}. Each spot shows a speckled ring and no signs of crystallinity. Detector defects are highlighted by a box throughout the figure. \textbf{(c)} A 1D intensity cut, $I(k)$, for 8 different regions as well as the average intensity. A peak is observed $\sim\SI{3.2}{nm^{-1}}$. Detector defects are highlighted by a box. \textbf{(d)} FEM variance, $V(k)$, as a function of scattering vector $k$ for amorphous (x10 to enable comparison since the magnitude of $V(k)$ is so much smaller) and nanocrystalline Bi$_2$Se$_3$. The nanocrystalline sample exhibits substantial variation in intensity for a given $k$-vector from crystalline Bragg diffraction peaks leading to a large variance, while the amorphous sample shows little variation. Detector defects are highlighted by a box. \textbf{(e)} A XRD $2\theta$ scan for amorphous Bi$_2$Se$_3$ after the Se decap showing the same broad low angle peak near 17$\degree$ and no signs of incipient crystallization. XRD provides a macroscopic probe of the films structure. The substrate peak is labeled. \textbf{(f)} Raman spectra for \SI{50}{nm} amorphous Bi$_2$Se$_3$ films using a \SI{488}{nm} laser. The peaks are labeled with their respective Raman mode. Different curves (blue and purple) correspond to different laser powers, showing the bulk Raman modes become more well-defined and do not shift. All samples presented in this work show similar spectra to the lower power spectra shown in this figure. Crystalline data \cite{PhysRevB.98.045411} (green curve) is overlayed to show the lack of a Van der Waal mode at $\sim$\SI{72}{cm^{-1}} in the amorphous films and the extra peak in the amorphous film at $\sim$\SI{250}{cm^{-1}} which is not seen in crystalline Bi$_2$Se$_3$, associated with Se-Se bonding.

\clearpage

\begin{figure*}
\centering
  \includegraphics[width=1\textwidth]{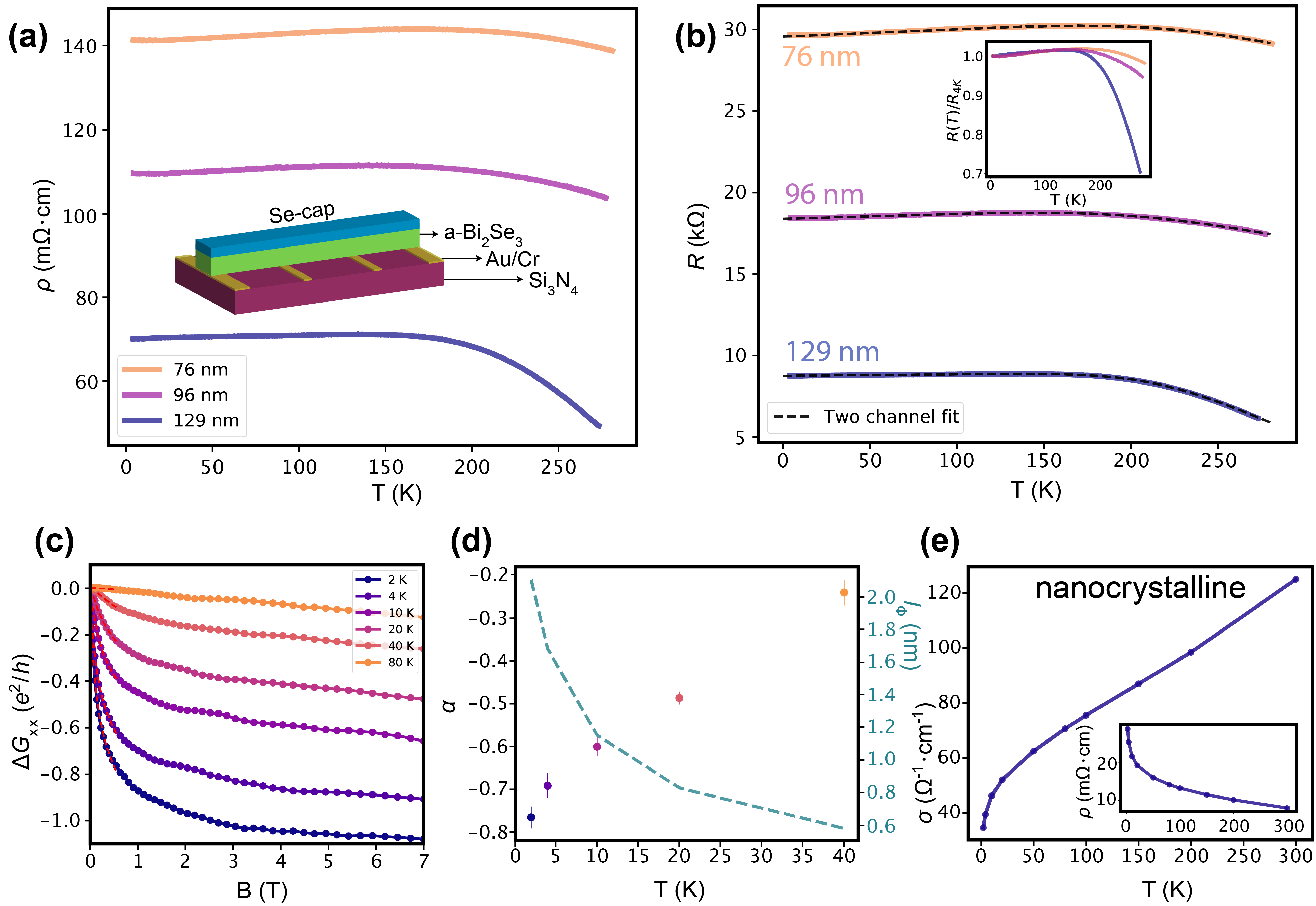}
\end{figure*}
\noindent {\bf Fig. 2.} \textbf{Electron transport in amorphous Bi$_2$Se$_3$} \textbf{(a)} $\rho\left(T\right)$ for \SIlist{76;96;129}{nm} films. All films show a high resistivity with little temperature dependence. Inset: Schematic of the structure used to measure resitivity. \textbf{(b)} The resistance for \SIlist{76;96;129}{nm} films. All films show high temperature VRH behavior (inset) and low temperature metallic behavior in $R$ with a low temperature saturation. Two-channel conductance fits the data reasonably well indicating a metallic surface and insulating VRH bulk behavior. \textbf{(c)} Conductance change as a function of the magnetic field for a \SI{140}{nm} film, measured at \SIlist{2;4;10;20;40;80}{K}, where $\Delta G_{xx} = G\left(B\right)-G\left(0\right)$. The deep cusp in the low field regime is characteristic of the WAL effect. \textbf{(d)} Magnetoconductance HLN fits showing $\alpha$ values indicating decoupled surface surface states at \SI{2}{K} and a single conduction channel at \SI{20}{K}. The dephasing length $l_{\phi}$ decreases with increasing temperature. \textbf{(e)} Nanocrystalline Bi$_2$Se$_3$ conductivity as a function of temperature. The conductivity drops with decreasing temperature. Resistivity is shown in the inset.

\clearpage

\begin{figure}[ht!]
\centering
  \includegraphics[width=1\textwidth]{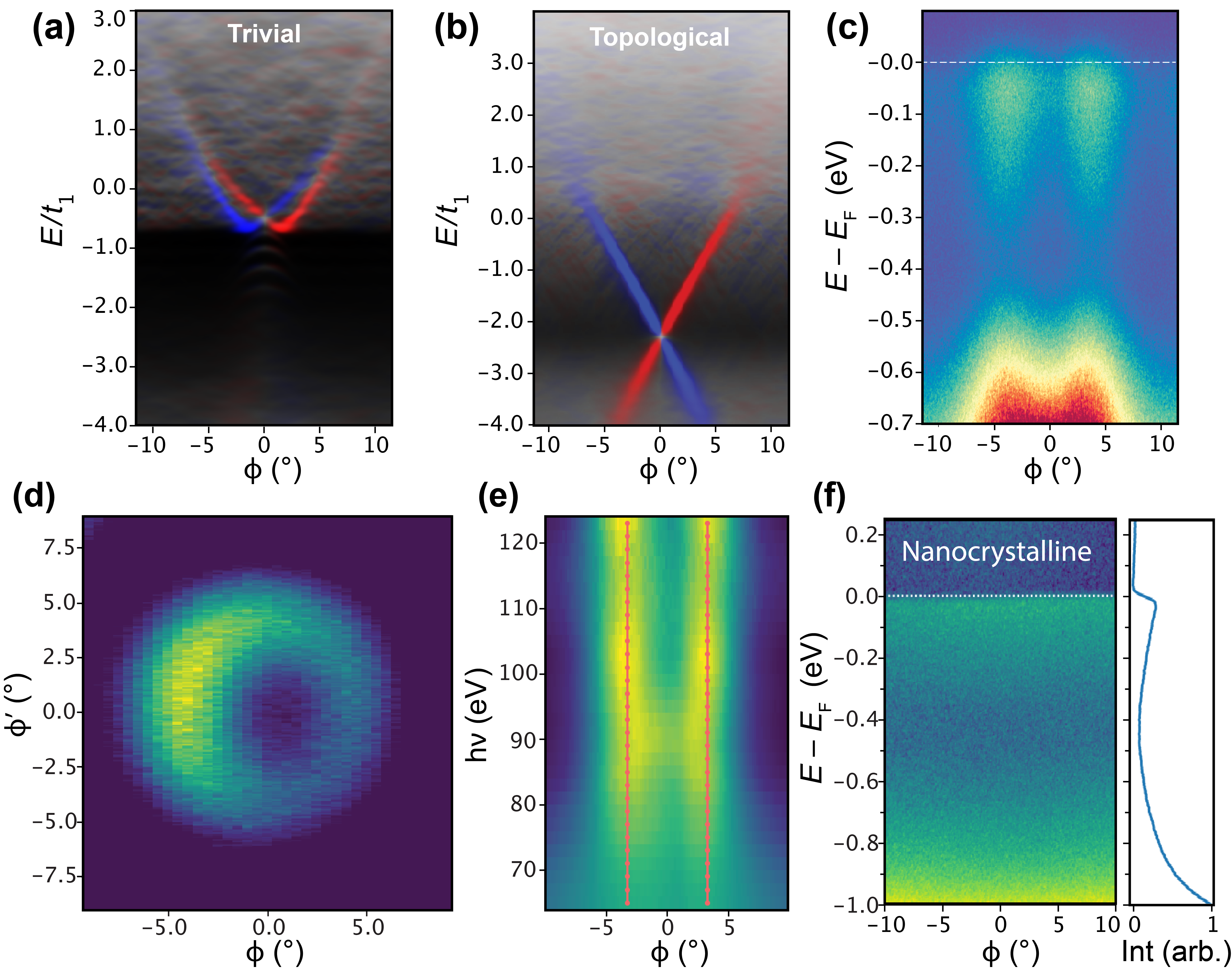}
\end{figure}
\noindent {\bf Fig. 3.} \textbf{ARPES spectra of electronic states in amorphous Bi$_2$Se$_3$} A calculated spin-resolved surface spectral function as a function of $\phi$ for the \textbf{(a)} trivial and \textbf{(b)} topological phase. In the topological phase the Dirac point is low in binding energies and Rashba spin-split states develop near the Fermi level. \textbf{(c)} ARPES spectrum $E$ vs. $\phi$ taken at normal emission at $h\nu=\SI{117.5}{eV}$. The spectrum reveals vertical states that cross the bulk gap and meet at \SI{-0.6}{eV} near the bulk valence states. \textbf{(d)} The ring-like in-plane Fermi surface. $\phi$ are the angles simultaneously collected by the detector referenced to normal incidence at a given sample tilt $\phi'$. \textbf{(e)} $h\nu$ vs. $\phi$ with binding energy integrated from \SI{-0.6}{eV} to the Fermi level and normalized by photon energy. The $h\nu$ vs. $\phi$ plot displays no photon energy dependence of the photoemission angle. Red dotted lines are fit to intensity peaks in the $h\nu$ vs. $\phi$ spectrum. \textbf{(f)} ARPES spectrum $E$ vs. $\phi$ for nanocrystalline Bi$_2$Se$_3$ showing an obvious lack of dispersion.

\clearpage

\begin{figure}
\centering
  \includegraphics[width=1\textwidth]{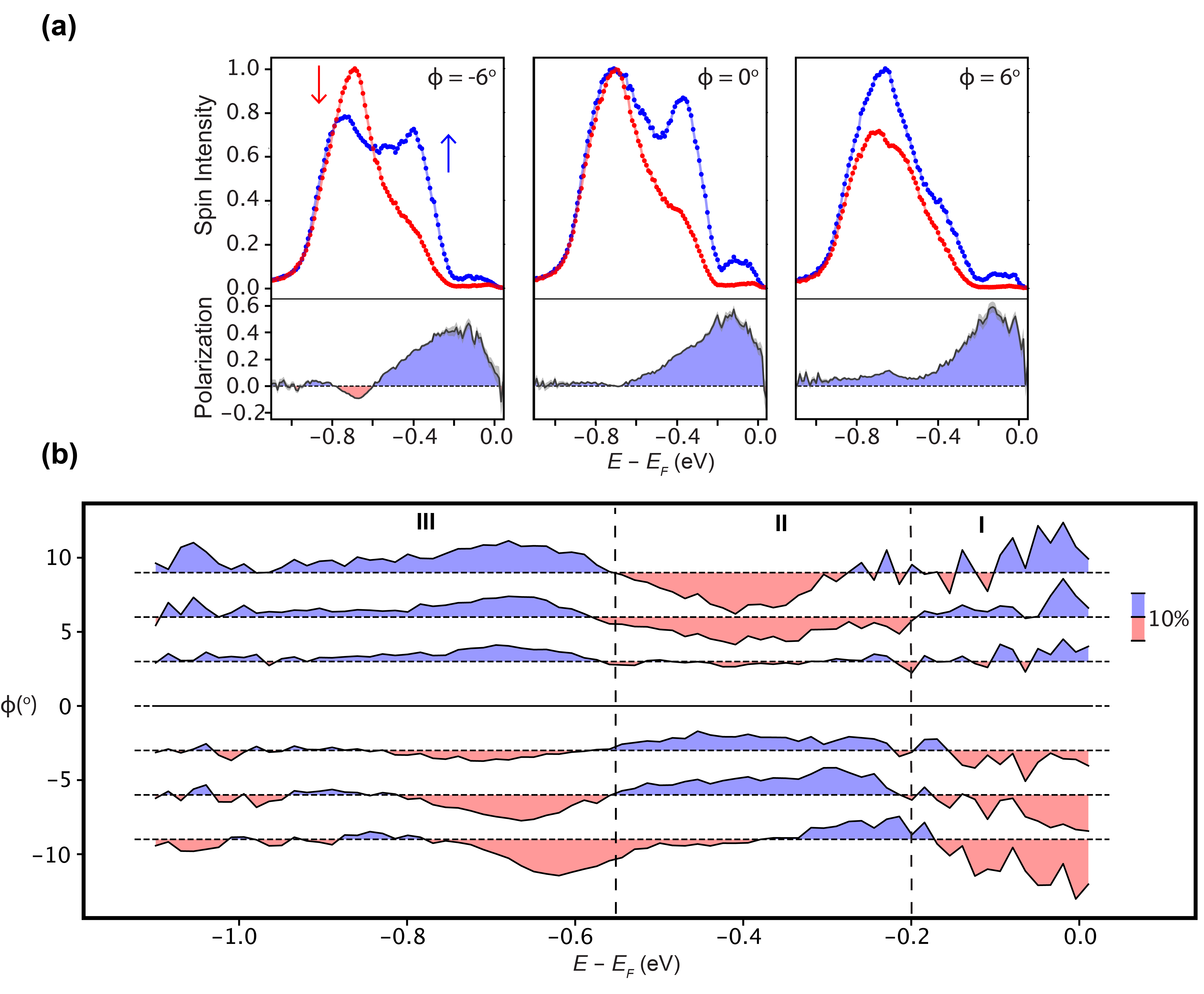}
\end{figure}
\noindent {\bf Fig. 4.} \textbf{Spin-resolved ARPES spectra of electronic states in amorphous Bi$_2$Se$_3$} \textbf{(a)} Spin-resolved EDC's taken at $\phi=\SI{-6}{\degree}$, $\Gamma$, and $\phi=\SI{6}{\degree}$, respectively. The spin contributions at each binding energy vary with respect to $\phi=\SI{0}{\degree}$. \textbf{(b)} Spin-resolved EDC map of $E$ vs $\phi$ with SME background subtraction taken from $\phi=\SI{-9}{\degree}$ to $\phi=\SI{9}{\degree}$. The spin polarization switches from red to blue (or vice versa on the other side of $\Gamma$) at \SI{-0.2}{eV} and from blue to red at \SI{-0.55}{eV}.

\end{document}